\documentclass{article}

\usepackage{subfigure}
\usepackage{spconf}
\usepackage{amsmath,amssymb}
\usepackage[dvips]{graphicx}
\usepackage{amsfonts}
\usepackage[mathscr]{eucal}
\usepackage{latexsym}
\usepackage{amsthm}
\usepackage{exscale}
\usepackage[mathscr]{eucal}
\usepackage{bm}
\usepackage[dvipsnames]{color}
\usepackage{cases}
\usepackage{epsfig}
\usepackage[center,small]{caption}
\usepackage{algorithm}
\usepackage{algorithmic}
\usepackage[verbose,nospace,sort]{cite}
\usepackage{tabularx}
\usepackage{multirow}
\usepackage{multicol}
\usepackage{balance}

\graphicspath{{./figs/}}

\scrollmode

\newcommand{\hSDsq}{|h_{SD}|^2}
\newcommand{\hSEsq}{|h_{SE}|^2}
\newcommand{\hEDsq}{|h_{ED}|^2}
\newcommand{\gammatd}{\tilde{\gamma}}
\newcommand{\lk}{\mathrm{leak}}
\newcommand{\Ptd}{\tilde{P}}

\begin{document}
\title{Active Eavesdropping via Spoofing Relay Attack \vspace{-1ex}}
\name{Yong~Zeng and Rui~Zhang
\address{Department of Electrical and Computer Engineering, National University of Singapore\\
\{elezeng, elezhang\}@nus.edu.sg \vspace{-2ex}}
}
\maketitle

\begin{abstract}
This paper studies a new active eavesdropping technique  via the so-called {\it spoofing relay attack}, which could be launched by the eavesdropper to significantly enhance the {\it information leakage rate} from the source over conventional passive eavesdropping. With this attack, the eavesdropper acts as a relay to spoof the source to vary transmission rate in favor of its eavesdropping performance by either enhancing or degrading the effective channel of the legitimate link. The maximum information leakage rate achievable by the eavesdropper and the corresponding optimal operation at the spoofing relay are obtained. It is shown that such a spoofing relay attack could impose new challenges from a physical-layer security perspective since it leads to significantly higher information leakage rate than conventional passive eavesdropping.
\end{abstract}

\section{Introduction}
Wireless communications are vulnerable to eavesdropping by unintended recipients due to the broadcast nature of wireless channels. The conventional cryptographic mechanism~\cite{603}, though provides an effective approach for secure communications, is facing with unprecedented challenges due to the fast growing computation power of the eavesdroppers, the increased complexity in  key generation and management, etc. Recently, there has been a significant research interest in achieving secure wireless communications by exploiting the inherent  wireless channel characteristics of the legitimate and adversary users, which is known as {\it physical-layer security} \cite{596}. Under the classic wiretap channel framework~\cite{599}, numerous efforts have been devoted to characterizing the {\it secrecy capacity} \cite{601,607,602}, or the maximum transmission rate at which the message can be reliably decoded at the legitimate receiver without leaking any useful information to the eavesdropper.

 Most of the existing works on physical-layer security have assumed the theoretical setup with passive eavesdroppers only. In practice, the eavesdropper could launch proactive attacks to enhance their eavesdropping performance, a technique known as {\it active eavesdropping} \cite{608}. For instance, in multi-antenna time-division duplexing (TDD) systems with reverse-link channel training, the eavesdropper may attack the channel training phase by sending identical pilots as the legitimate receiver, so that the estimated channel at the source transmitter, based on which precoding is designed for the data transmission phase, is a linear combination of those of the legitimate and eavesdropping links.  Such an active attack is known as {\it pilot contamination attack} \cite{604}, by which the eavesdropper can enhance its effective channel from the source transmitter, and hence boost its eavesdropping capacity, while simultaneously degrading the channel of the legitimate link. Various schemes have been proposed  to detect such a pilot contamination attack \cite{605,609,606,610,611}.

In this paper, we study a new active attack termed {\it spoofing relay attack}, which could be launched by the eavesdropper to significantly enhance the effective {\it information leakage rate} eavesdropped from the source over the conventional passive eavesdropping. With this attack, the eavesdropper acts as a relay to spoof the source to vary transmission rate in favor of its eavesdropping performance,  assuming that adaptive rate transmission is adopted at the source based on the effective channel to the legitimate receiver. Specifically, if the eavesdropper has a better channel than that of the legitimate receiver, it will enhance the effective channel of the legitimate link by forwarding a constructive signal to the receiver, which leads to higher transmission rate by the source, and hence higher information leakage rate; otherwise, it will degrade the effective channel of the legitimate link via forwarding a destructive signal to the receiver, so as to spoof the source to reduce transmission rate to make it decodable by the eavesdropper. The maximum information leakage rate achievable by such a spoofing relay attack is derived, which is shown to be significantly higher than that attainable by conventional passive eavesdropping.

Compared to other active eavesdropping techniques such as the pilot contamination attack, the spoofing relay attack could lead to more severe security risks, since it has a broader applicability, regardless of single- or multi-antenna, TDD or frequency-division duplexing (FDD) systems. Furthermore, it
is also more difficult to be detected, since the legitimate user may attribute the change in its effective channel to the environmental variations, e.g., the presence of a new signal path. Devising effective detection schemes and countermeasures against the new spoofing relay attack is an interesting problem, which is left for our future work.

\begin{figure}
\centering
\includegraphics[scale=0.7]{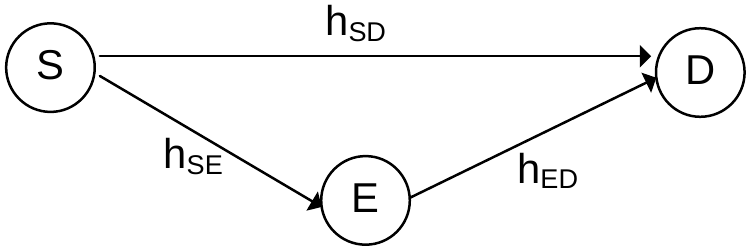}
\caption{A point-to-point link with an active eavesdropper.\vspace{-3ex}}\label{F:model}
\end{figure}

\section{System Model and Problem Formulation}
As shown in Fig.~\ref{F:model}, we consider a point-to-point wireless communication system where the source  $\mathbf S$  sends information to the destination $\mathbf D$ in the presence of an eavesdropper $\mathbf E$. We assume that {\it adaptive rate transmission} is adopted at $\mathbf S$ based on the channel condition perceived at $\mathbf D$. However, both $\mathbf S$ and $\mathbf D$ are unaware of the presence of $\mathbf E$, so that no dedicated coding as in conventional physical-layer security  (see e.g. \cite{596}-\cite{602}) is applied to prevent the eavesdropping by $\mathbf E$. On the other hand, the eavesdropper $\mathbf E$ can conduct either passive or active eavesdropping, as discussed below.
\vspace{-2ex}
\subsection{Passive Eavesdropping}\label{sec:passive}
With passive eavesdropping, $\mathbf E$ remains silence throughout the communication between $\mathbf S$ and $\mathbf D$, but tries to decode the information from $\mathbf S$. In this case, the channel capacity of the legitimate link from $\mathbf S$ to $\mathbf D$, which is also  assumed to be the transmission rate by $\mathbf S$, is
$R_D=\log_2\left(1+P_S|h_{SD}|^2/\sigma^2 \right)$ in bits/second/Hz (bps/Hz),
where $h_{SD}$ is the complex-valued channel gain from $\mathbf S$ to $\mathbf D$, $P_S$ is the transmission power at $\mathbf S$, and $\sigma^2$ is the power of the additive white Gaussian noise (AWGN) at $\mathbf D$. Similarly, the channel capacity between $\mathbf S$ and $\mathbf E$ is
$R_E=\log_2\left(1+P_S|h_{SE}|^2/\sigma^2\right)$ in bps/Hz,
with $h_{SE}$ denoting the channel from $\mathbf S$ to $\mathbf E$. If $R_E\geq R_D$ or equivalently $|h_{SE}|^2\geq |h_{SD}|^2$, i.e., the eavesdropper has a better channel than the legitimate receiver, $\mathbf E$ can reliably decode the information sent by $\mathbf S$ with arbitrarily small error. As a result, the effective {\it information leakage rate} is given by $R_{\lk}=R_D$. On the other hand, if $R_E<R_D$, or the eavesdropper has a weaker channel than the legitimate receiver, then it is impossible for $\mathbf E$ to decode the information from $\mathbf S$ with arbitrarily small error. In this case, we define the effective information leakage rate as $R_{\lk}=0$.\footnote{Note that in this case $\mathbf E$ may still extract useful information from its received signal; while in this paper we consider a more stringent setup where the message from $\mathbf S$ needs to be decoded at $\mathbf E$ with arbitrarily small error.} Therefore, the information leakage rate can be expressed as
\begin{align} \label{eq:Rlk}
R_{\lk}=\begin{cases}
R_D, & \text{if } R_E\geq R_D \\
0, & \text{otherwise}.
\end{cases}
\end{align}

\vspace{-2ex}
\subsection{Active Eavesdropping via Spoofing Relay Attack} 
\begin{figure}
\centering
\includegraphics[scale=0.6]{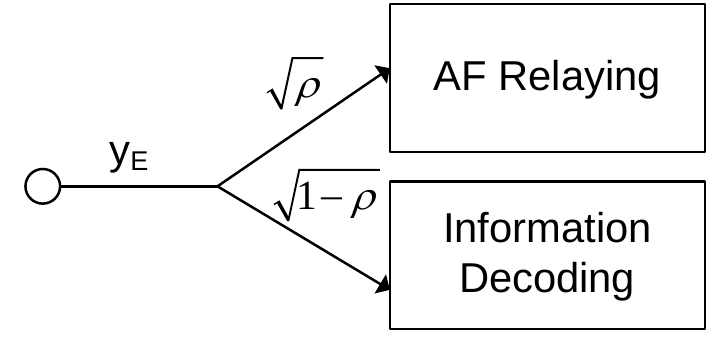}
\caption{The architecture of a spoofing relay.\vspace{-3ex}}\label{F:architecture}
\end{figure}

\vspace{-1ex}
In this subsection, we consider an  active eavesdropper that launches the spoofing relay attack to  enhance the information leakage rate.
 With such an attack, the eavesdropper $\mathbf E$ operates in a full-duplex mode with simultaneous information reception and relaying \cite{612}. We assume the simple amplify-and-forward (AF) relaying by $\mathbf E$ since it incurs the minimal processing delay. By assuming an ideal full-duplex operation with perfect self-interference cancellation \cite{612}, the signal received by  $\mathbf E$ prior to processing noise addition is $y_E=h_{SE}\sqrt{P_S}d_S$, where $d_S\sim \mathcal{CN}(0,1)$ denotes the circularly-symmetric complex Gaussian (CSCG) distributed information-bearing symbol sent by $\mathbf S$. As shown in Fig.~\ref{F:architecture}, the received signal $y_E$ is split into two parts at $\mathbf E$, one for information relaying aiming to alter the effective channel of the legitimate link from $\mathbf S$ to $\mathbf D$, and the other for information decoding so as to eavesdrop the message sent by $\mathbf S$. Denote by $0\leq \rho \leq 1$ the power splitting ratio for the signal part split  for information relaying. The transmitted signal $x_E$ by $\mathbf E$ can then be expressed as
\begin{align}
x_E=v \left( \sqrt{\rho}h_{SE}\sqrt{P_S}d_S +n_E^{(R)} \right), \label{eq:xE}
\end{align}
where $v$ is the complex-valued amplification coefficient at $\mathbf E$, and $n_E^{(R)}\sim \mathcal{CN}(0,\sigma^2)$ denotes the AWGN introduced during the relaying operation at $\mathbf E$.
By assuming that the processing delay due to the AF relaying at  $\mathbf E$ is negligible, the signal received at $\mathbf D$ can be expressed as
\begin{align}
\hspace{-2ex}  y_D & =h_{SD}\sqrt{P_S}d_S+h_{ED}x_E+n_D,\\
 & \hspace{-2ex} = \left( h_{SD}+ v \sqrt{\rho} h_{SE} h_{ED}\right)  \sqrt{P_S}d_S
+ v  h_{ED} n_E^{(R)}+n_D, \label{eq:yD}
\end{align}
where $h_{ED}$ denotes the channel from $\mathbf E$ to $\mathbf D$, and $n_D\sim \mathcal{CN}(0,\sigma^2)$ is the AWGN at $\mathbf D$. It is observed from \eqref{eq:yD} that by adjusting the power splitting ratio $\rho$ and the amplification coefficient $v$, the eavesdropper $\mathbf E$ is able to alter the effective channel  from $\mathbf S$ to $\mathbf D$. 
The effective capacity of the legitimate link  can then be expressed as $\tilde{R}_{D} = \log_2(1+\tilde{\gamma}_D)$, where $\tilde{\gamma}_D$ is the effective signal-to-noise ratio (SNR) at $\mathbf D$, which can be obtained from \eqref{eq:yD} as a function of $\rho$ and $v$, given by
\begin{align}
\tilde{\gamma}_D(\rho, v)=\frac{\left| h_{SD}+v \sqrt{\rho}h_{SE}h_{ED}\right|^2 P_S}{(1+|v|^2 |h_{ED}|^2)\sigma^2}. \label{eq:gammatdD}
\end{align}

On the other hand, at the information decoder of $\mathbf E$, the signal based on which the message from $\mathbf S$ is decoded can be expressed as
\begin{align}
\tilde{y}_E &=\sqrt{1-\rho} h_{SE}\sqrt{P_S}d_S + n_E^{(D)},
\end{align}
where $n_E^{(D)}\sim \mathcal{CN}(0,\sigma^2)$ denotes the AWGN at the information decoder of $\mathbf E$. Thus, the  information rate achievable by $\mathbf E$ is $\tilde{R}_{E}=\log_2(1+\tilde{\gamma}_E)$, where $\tilde{\gamma}_E$ is the  SNR as a function of $\rho$ given by
\begin{align}
\tilde{\gamma}_E(\rho)=\frac{(1-\rho)|h_{SE}|^2P_S}{\sigma^2}.
\end{align}

To study the worst-case scenario under the spoofing relay attack, we assume that perfect channel state information (CSI) of all links is available at $\mathbf E$. The investigation on the  spoofing relay attack with imperfect or limited CSI at $\mathbf E$ is left for our future work. The objective of $\mathbf E$ is to optimize the power splitting ratio $\rho$ and the amplification coefficient $v$ so that the information leakage rate is maximized. Based on the definition in \eqref{eq:Rlk}, the problem can be formulated as
\begin{align}
\mathrm{(P1)}:
\begin{cases}
  \underset{v, \rho}{\max}  &   \ \tilde{R}_D   \\
\text{ s.t. }  &  \tilde{R}_E \geq \tilde{R}_D \\
& 0\leq \rho \leq 1, \\
& |v|^2  \left( \rho |h_{SE}|^2P_S+\sigma^2\right) \leq P_E,
\end{cases}
\end{align}
where  $P_E$ denotes the maximum transmission power at $\mathbf E$.

\vspace{-1ex}
\section{Optimal Solution \vspace{-1ex}}\label{sec:optSol}
To find the optimal solution to $\mathrm{(P1)}$, notice that $\tilde{R}_D$ and $\tilde{R}_E$ in $\mathrm{(P1)}$ can be respectively replaced by $\gammatd_D(v,\rho)$ and $\gammatd_E(\rho)$ due to their monotonic relations. Furthermore, for any fixed power splitting ratio $0\leq \rho\leq 1$, we first obtain the maximum achievable SNR at $\mathbf D$, denoted as $\gammatd_D^{\max}(\rho)$, by optimizing the amplification coefficient $v$ as
\begin{align}
\gammatd^{\max}_D(\rho)\triangleq
\begin{cases}\label{eq:gammaDMax}
 \underset{v}{\max } &\  \gammatd_D(\rho, v) \\
\text{s.t.} & |v|^2 \leq \frac{P_E}{\rho |h_{SE}|^2P_S+\sigma^2}.
\end{cases}
\end{align}
It follows from \eqref{eq:gammatdD} that at the optimal solution to \eqref{eq:gammaDMax}, the phase of $v$ should be chosen such that the two signal paths from $\mathbf S$ to $\mathbf D$ add constructively, i.e., $\angle v =\angle h_{SD}-\angle h_{SE} - \angle h_{ED}$, where $\angle z$ denotes the phase of a complex number $z$. We term such a strategy of the spoofing relay as {\it constructive information forwarding}, since it helps enhance the effective channel of the legitimate link from $\mathbf S$ to $\mathbf D$. In addition, the magnitude of the optimal $v$ to \eqref{eq:gammaDMax} can be obtained by examining its first-order derivative, and the resulted maximum SNR can be expressed as
\begin{align}
\gammatd_{D}^{\max}(\rho)=
\begin{cases}\notag
 \left(1+\frac{\rho |h_{SE}|^2}{|h_{SD}|^2} \right)\Ptd_S|h_{SD}|^2,  \ \hspace{6ex}  0\leq \rho \leq \rho_1 & \\
 \frac{\left(\sqrt{1+\rho \hSEsq\Ptd_S} +\frac{|h_{SE}||h_{ED}|}{|h_{SD}|}\sqrt{\rho \Ptd_E}\right)^2 \Ptd_S\hSDsq}{1+\rho \hSEsq\Ptd_S+\hEDsq\Ptd_E}, &  \\
 & \hspace{-12ex} \rho_1 < \rho \leq 1 ,
\end{cases}
\end{align}
where $\rho_1\triangleq \min \left\{1, \frac{-1+\sqrt{1+4\Ptd_S\Ptd_E|h_{SD}|^2 |h_{ED}|^2}}{2|h_{SE}|^2\Ptd_S}\right\}$, with $\Ptd_S\triangleq P_S/\sigma^2$ and $\Ptd_E\triangleq P_E/\sigma^2$. It can be verified that $\gammatd_{D}^{\max}(\rho)$ is a monotonically increasing function of $0\leq \rho \leq 1$. In particular, if $\rho=0$, i.e., no information forwarding is applied at $\mathbf E$, we have $v=0$ and $\gammatd_D^{\max}(0)=\Ptd_S|h_{SD}|^2$. This corresponds to the special case of passive eavesdropping previously discussed in Section~\ref{sec:passive}.

\begin{figure*}%
\centering
\subfigure[Case 1: $\hSDsq < \hSEsq$]{
\includegraphics[scale=0.43]{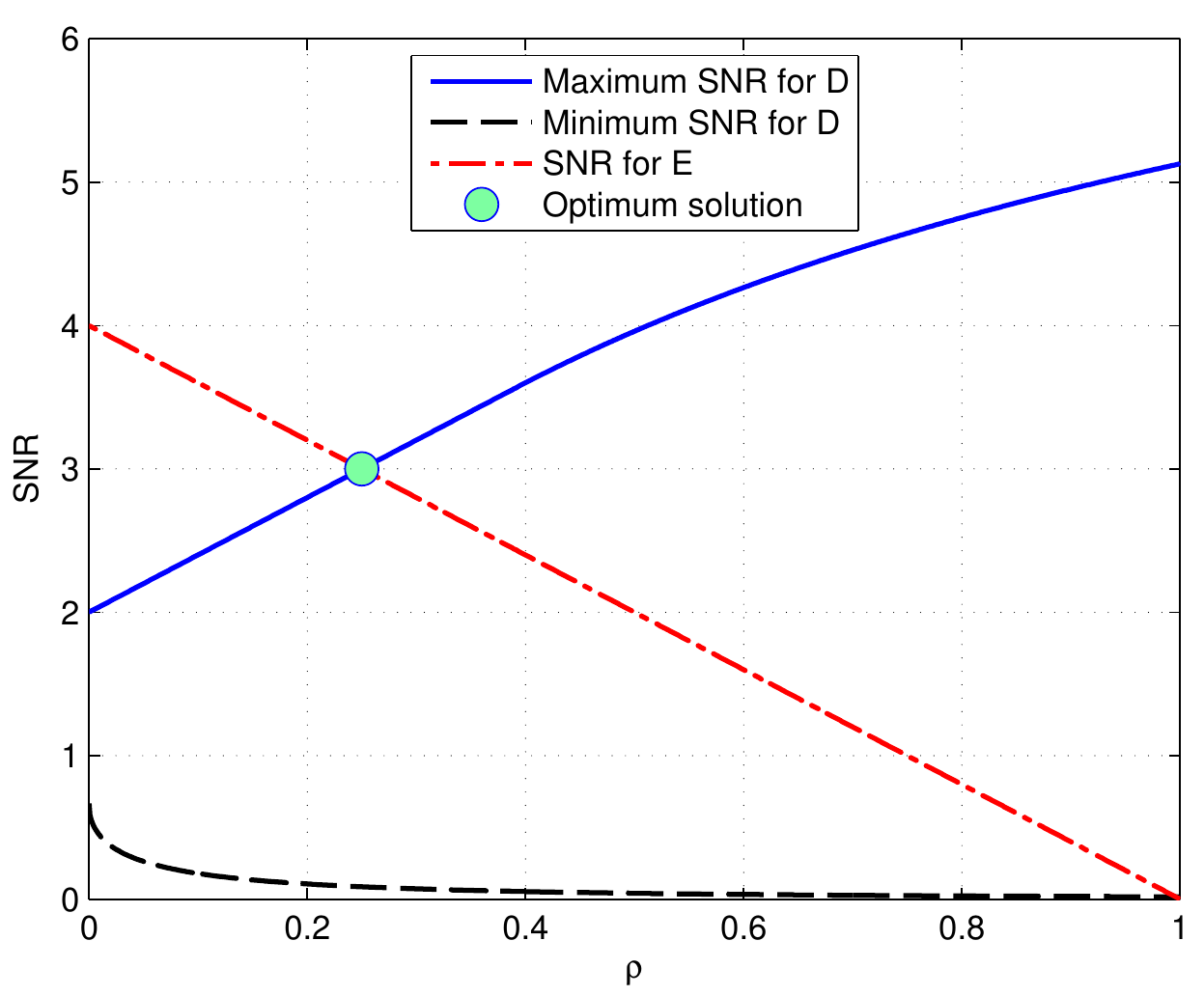}
\label{fig:subfigure1}
}
\subfigure[Case 2: $\frac{\hSDsq}{1+\hEDsq\Ptd_E}\leq \hSEsq \leq \hSDsq$]{
\includegraphics[scale=0.43]{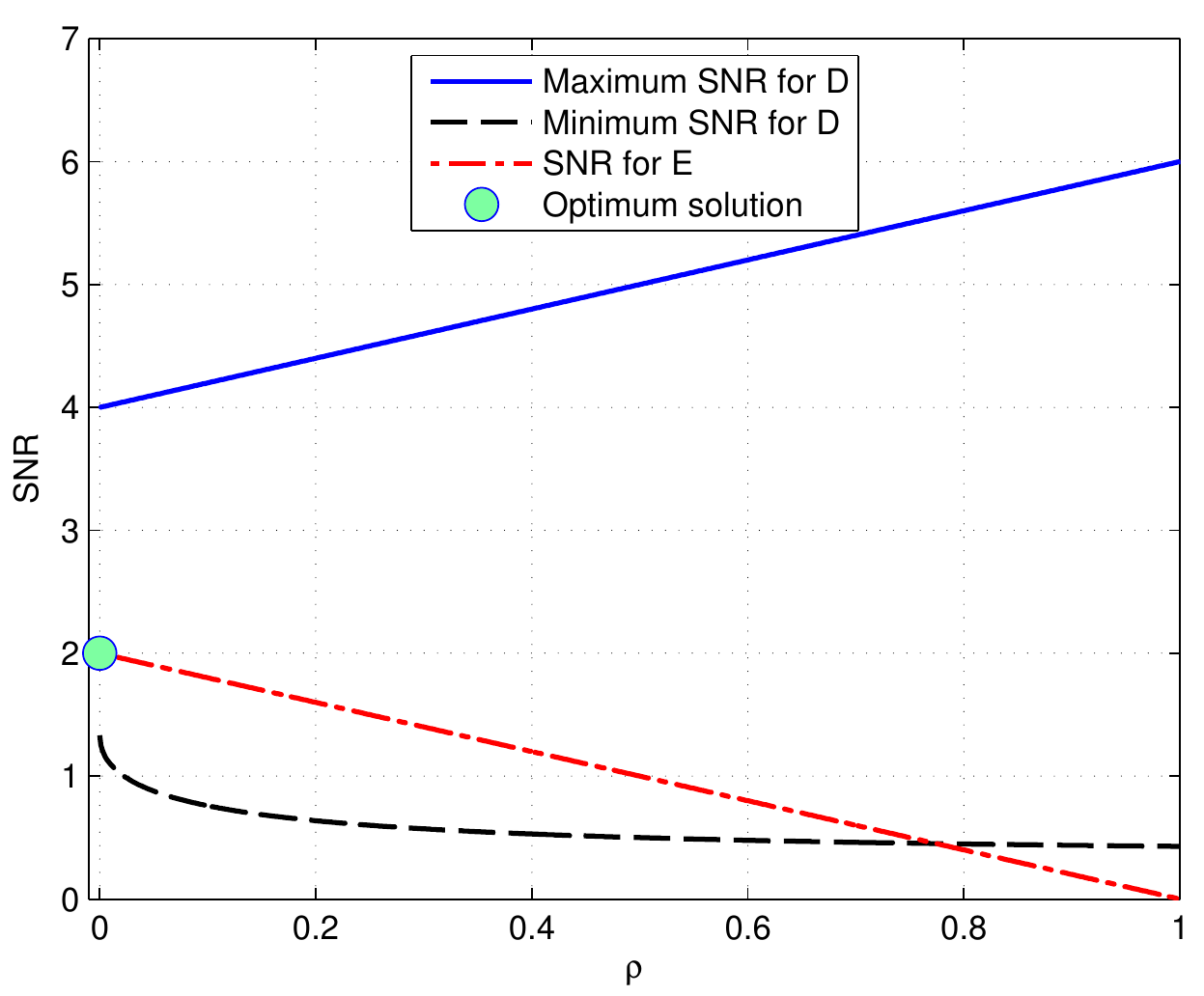}
\label{fig:subfigure2}
}
\subfigure[Case 3: $\hSEsq < \frac{\hSDsq}{1+\hEDsq\Ptd_E}$]{
\includegraphics[scale=0.43]{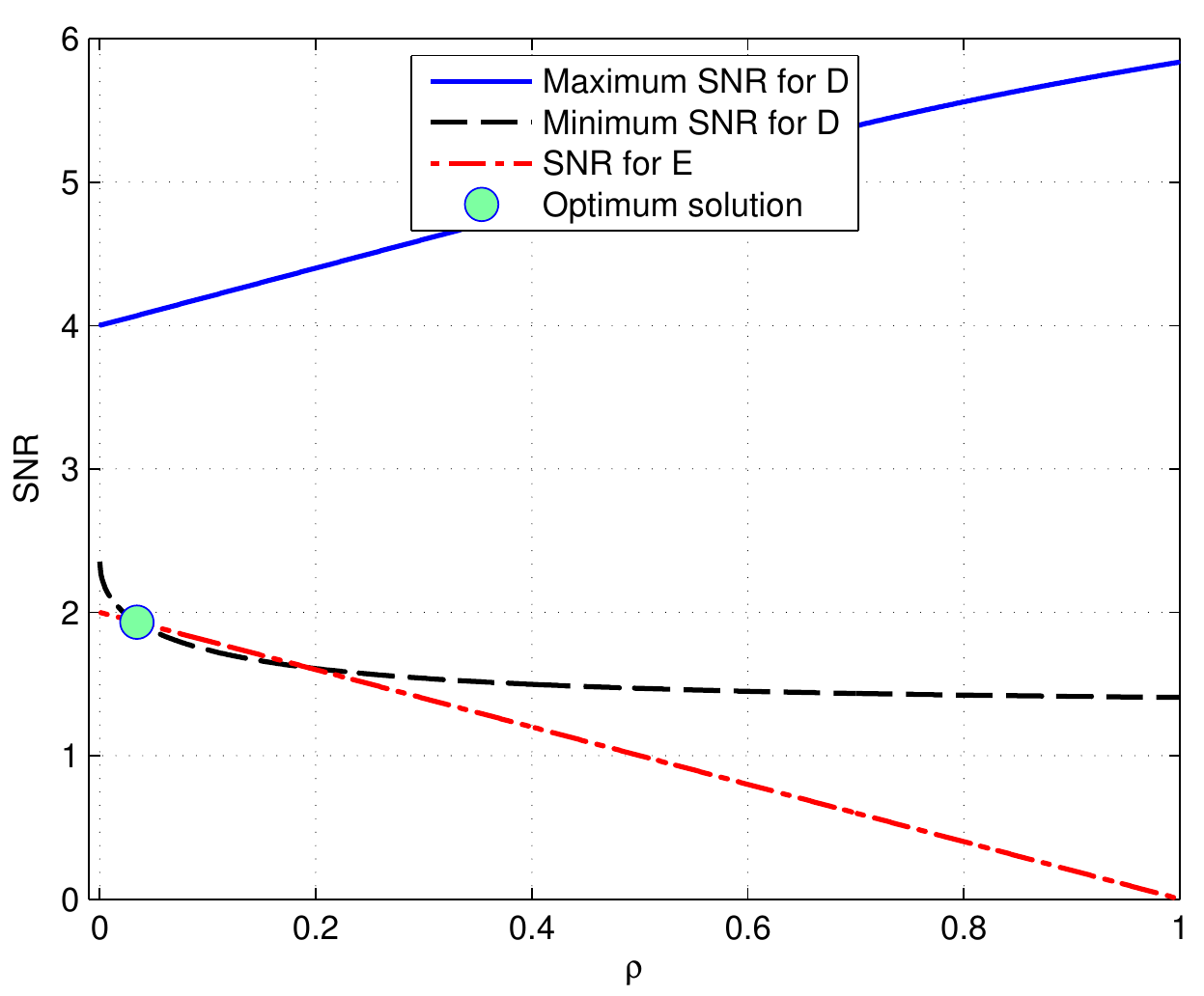}
\label{fig:subfigure3}
}
\vspace{-3ex}
\caption{Three cases for the optimal power splitting solution.\vspace{-3ex}}%
\label{F:antennas3D}
\end{figure*}

On the other hand, for fixed $0\leq \rho \leq 1$, the minimum achievable SNR at $\mathbf D$, denoted as $\gammatd_D^{\min}(\rho)$, can be obtained by solving
\begin{align}
\gammatd^{\min}_D(\rho)\triangleq
\begin{cases}\label{eq:gammaDMin}
 \underset{v}{\min } &\  \gammatd_D(\rho, v) \\
\text{s.t.} & |v|^2 \leq \frac{P_E}{\rho |h_{SE}|^2P_S+\sigma^2}.
\end{cases}
\end{align}
It follows from \eqref{eq:gammatdD} that at the optimal solution to \eqref{eq:gammaDMin}, the two signal paths from $\mathbf S$ to $\mathbf D$ should add destructively, i.e., $\angle v = \pi + \angle h_{SD}-\angle h_{SE} - \angle h_{ED}$. Such a strategy at $\mathbf E$ is termed as {\it destructive information forwarding}, which essentially degrades the effective channel of the legitimate link from $\mathbf S$ to $\mathbf D$. Furthermore, by taking the first order derivative with respect to the magnitude of $v$, the corresponding optimal value of \eqref{eq:gammaDMin} can be expressed as
\begin{align}
\gammatd_{D}^{\min}(\rho)=
\begin{cases}\notag
 \frac{\left( \sqrt{1+\rho \hSEsq\Ptd_S} -\frac{|h_{SE}||h_{ED}|}{|h_{SD}|}\sqrt{\rho \Ptd_E} \right)^2\Ptd_S\hSDsq }{1+\rho \hSEsq\Ptd_S+\hEDsq\Ptd_E}, & \\
& \hspace{-13ex} 0 \leq \rho \leq \rho_2 \\
 0,   &  \hspace{-13ex} \rho_2 < \rho \leq 1 ,
\end{cases}
\end{align}
where $\rho_2=C$ if $0\leq C \leq 1$, and $\rho_2=1$ otherwise, with $C\triangleq \frac{\hSDsq}{\hSEsq\left(\hEDsq\Ptd_E-\hSDsq\Ptd_S\right)}$. In particular, if $\rho=0$, i.e., no information forwarding by $\mathbf E$, we have $\gammatd_{D}^{\min}(0)=\frac{\Ptd_S\hSDsq}{1+\hEDsq\Ptd_E}$. This corresponds to degrading the SNR at $\mathbf D$  via {\it jamming}, i.e., by amplifying the noise with full power  at $\mathbf E$. For $\rho>0$, both destructive information forwarding and jamming (i.e., noise amplification) contribute to the SNR degradation at $\mathbf D$, as can be seen from the expression of $\gammatd_{D}^{\min}(\rho)$.

Since $\gammatd_{D}(\rho, v)$ is a continuous function of $v$, for any fixed $0\leq \rho \leq 1$, the set of  achievable SNRs at $\mathbf D$ is given by the interval $\left[\gammatd_{D}^{\min}(\rho),  \gammatd_{D}^{\max}(\rho)\right]$. Consequently, $\mathrm{(P1)}$ reduces to finding the optimal power splitting ratio $\rho$ via solving
\begin{align}
\mathrm{(P2)}:
\begin{cases} \underset{0\leq \rho \leq 1} {\max} & \ \gammatd_D(\rho)   \\
\text{ s.t. } & \gammatd_{D}^{\min}(\rho) \leq \gammatd_D(\rho) \leq \gammatd_{D}^{\max}(\rho) \\
& \gammatd_D(\rho)\leq \gammatd_E(\rho),
\end{cases}
\end{align}
which can be solved by considering the following three cases.

{\it Case 1:} $\gammatd_D^{\max}(0)<\gammatd_E(0)$, or $\hSDsq < \hSEsq$, as illustrated in Fig.~\ref{fig:subfigure1}. In this case, $\mathbf E$ has a better channel than the legitimate receiver $\mathbf D$. Intuitively, $\mathbf E$ should perform constructive information forwarding to enhance the effective channel of $\mathbf D$ so as to increase the information leakage rate. It follows from Fig.~\ref{fig:subfigure1} that the optimal solution to $\mathrm{(P2)}$ is given by the intersection point of the two curves $\gammatd_D^{\max}(\rho)$ and $\gammatd_E(\rho)$. As $\gammatd_D^{\max}(\rho)$ and $\gammatd_E(\rho)$  are monotonically increasing and decreasing functions over $0\leq \rho \leq 1$, respectively, and $\gammatd_D^{\max}(1)>\gammatd_E(1)=0$, the equation $\gammatd_D^{\max}(\rho)=\gammatd_E(\rho)$ has one unique solution $\rho^\star$, which can be obtained numerically.


{\it Case 2:} $\gammatd_D^{\min}(0)\leq \gammatd_E(0)\leq \gammatd_D^{\max}(0)$, or $\frac{\hSDsq}{1+\hEDsq\Ptd_E}\leq \hSEsq \leq \hSDsq$, as illustrated in Fig.~\ref{fig:subfigure2}. In this case, the eavesdropping link is worse than the legitimate link, but it becomes better if jamming with full power is applied at $\mathbf E$ to degrade the legitimate link. It follows from Fig.~\ref{fig:subfigure2} that the optimal solution to $\mathrm{(P2)}$ is $\rho^\star=0$, i.e., no information forwarding and only  jamming is applied by $\mathbf E$ with normalized jamming power $\Ptd_E^\star=\frac{1}{\hEDsq}\left(\frac{\hSDsq}{\hSEsq}-1 \right)$ to degrade the legitimate link SNR to the same level as that at $\mathbf E$.

{\it Case 3:} $\gammatd_E(0)<\gammatd_D^{\min}(0)$, or $\hSEsq < \frac{\hSDsq}{1+\hEDsq\Ptd_E}$, as illustrated in Fig.~\ref{fig:subfigure3}. In this case, the eavesdropping link is worse than the legitimate link even after jamming with full power by $\mathbf E$. Therefore, destructive information forwarding and jamming should be both applied at $\mathbf E$ to further degrade the legitimate link.  It follows from Fig.~\ref{fig:subfigure3} that the optimal solution $\rho^\star$ to $\mathrm{(P2)}$ is obtained by solving $\gammatd_D^{\min}(\rho)=\gammatd_E(\rho)$ in the interval $0\leq \rho \leq 1$, which can be reduced to a quartic equation and hence solved efficiently. Note that if more than one solutions exist, the one with the smallest magnitude is the optimal solution. On the other hand, if no such a solution exists, it implies that problem $\mathrm{(P2)}$, and hence $\mathrm{(P1)}$, is infeasible, i.e., the spoofing relay attack is not sufficient to degrade the source transmission rate to a level achievable by the eavesdropper with its given power constraint.

\begin{figure}
\centering
\includegraphics[scale=0.24]{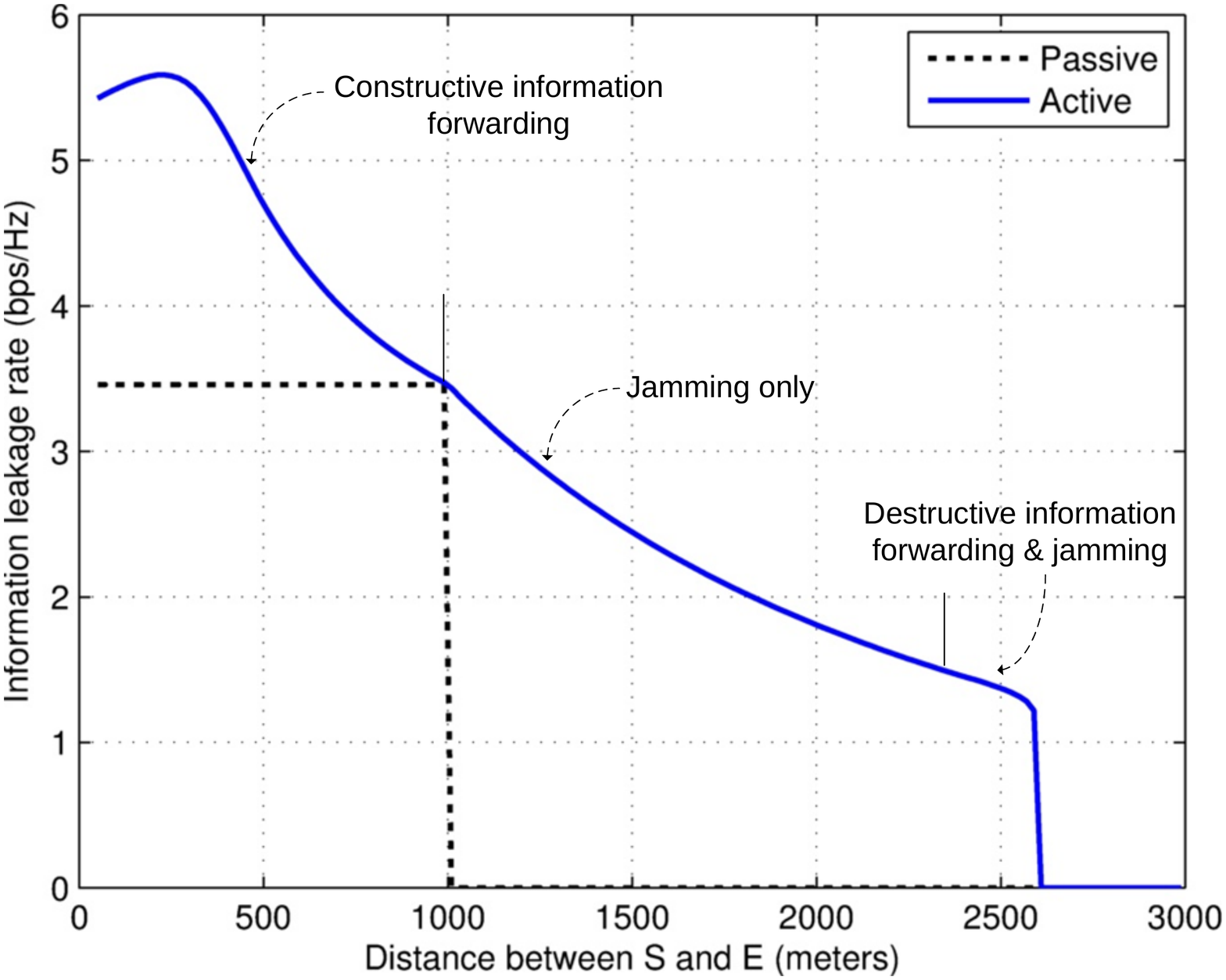}
\caption{The information leakage rate with passive versus active eavesdropping.\vspace{-3ex}}\label{F:RateVSDistance}
\end{figure}

\vspace{-2ex}
\section{Numerical Results \vspace{-1ex}}
We assume that the source $\mathbf S$ and the legitimate receiver $\mathbf D$  are separated by a fixed distance $d_{SD}=1000$ meters, and the eavesdropper $\mathbf E$ moves along the line from $\mathbf S$ to $\mathbf D$ with the distance $d_{SE}$ varying from $50$ to $3000$ meters. We assume line-of-sight (LoS) channels with free-space path loss model, and the operating frequency is assumed to be $1.8$ GHz. The source transmission power $P_S$ is set to a value such that the received SNR  at $\mathbf D$ (without eavesdropper's attack) is $10$ dB. By assuming $P_E=P_S$, Fig.~\ref{F:RateVSDistance} plots the information leakage rate $R_\lk$ versus $d_{SE}$  by  passive eavesdropping versus the studied active eavesdropping, with $R_\lk$ given by \eqref{eq:Rlk}. It is observed that with passive eavesdropping, a constant $R_\lk$, whose value is determined by the legitimate link, is achieved when $\mathbf E$ has a better channel than $\mathbf D$, i.e., $d_{SE}\leq d_{SD}$; whereas if $d_{SE}> d_{SD}$, $R_\lk$ drops to zero since $\mathbf E$ cannot reliably decode the information from $\mathbf S$. In contrast, with the active spoofing relay attack, $\mathbf E$ is able to achieve much higher information leakage rate. Fig.~\ref{F:RateVSDistance} also shows the three different strategies of the spoofing relay attack by the eavesdropper, namely constructive information forwarding, jamming, and both destructive information forwarding and jamming, which correspond to the three cases for determining the optimal power splitting ratio studied in Section~\ref{sec:optSol}.

\vspace{-2ex}\section{Conclusion}\vspace{-1ex}
This paper studies a new active eavesdropping technique via the spoofing relay attack. Depending on the channel conditions, the eavesdropper constructively or destructively forwards the information signal to the destination, so as to spoof the source to increase or decrease the transmission rate to maximize the information leakage rate. It is shown that with this new attack, the eavesdropper can significantly enhance the information leakage rate over the conventional passive eavesdropping. This paper opens a new avenue for investigating the physical-layer security with more intelligent eavesdroppers than conventional passive listeners.

\balance
\bibliographystyle{IEEEbib}
\bibliography{IEEEabrv,IEEEfull}

\end{document}